# Technique for Magnetic Susceptibility Determination in the High Doped Semiconductors by Electron Spin Resonance

A. I. Veinger, A. G. Zabrodskii, T. V. Tisnek, S. I. Goloshchapov, P. V. Semenikhin

Ioffe Institute of the Russian Academy of Sciences, St. Petersburg, Russia

Tel. +7 812 2927152

Fax +7 812 2976245

anatoly.veinger@mail.ioffe.ru

**Abstract.**

Method for determining the magnetic susceptibility in the high doped semiconductors is considered. A procedure that is based on double integration of the positive part of the derivative of the absorption line having a Dyson shape and takes into account the depth of the skin layer is described. Analysis is made for the example of arsenic-doped germanium samples at a rather high concentration corresponding to the insulator–metal phase transition.

## 1 INTRODUCTION

An increase in the impurity concentration in a semiconductor leads to a stronger interaction between impurity centers and to a second-order phase transition from the insulating to metallic state (IM), with conductivity appearing at zero temperature. The problem of phase transitions of this kind is widely discussed in the scientific literature [1, 2]. Together with the conductivity, the magnetic properties of the semiconductor change in the region of the IM transition. The Curie paramagnetism exhibited by noninteracting impurity electrons, characteristic of the insulating state, gives way to the Pauli paramagnetism of the metallic state. In contrast to the Curie paramagnetism, the Pauli paramagnetism is not contributed by all electrons. Thus, a study of the magnetic properties of a semiconductor near the IM phase transition involves estimation of the magnetic susceptibility. In principle, this can be done by means of the electron spin resonance (ESR) technique. However, a number of problems are encountered, associated with the pronounced change in the conductivity of a sample.

For example, in a lightly doped semiconductor at low temperatures, electrons are captured by impurity atoms (donors) and its resistivity is high. In ESR studies, a cavity with a sample of his kind has a high Q-factor. Finding the magnetic susceptibility $\chi_0$ in this case is a almost standard procedure described in detail in the literature [3, 4]. It is well known that the integral of the spin resonance absorption power $P$ is proportional to the magnetic susceptibility of the tested sample:

$$\chi_0 \propto \int P dH, \quad (1)$$

where $H$ is magnetic field.

However, an increase in conductivity gives rise to specific features that strongly complicate an analysis of ESR spectra. These problems are characteristic of any conducting crystal, irrespective of the nature of its conduction. Let us mention the arising problems in order in which they appear as the doping level is raised when the sample conductivity grows.

**Problem 1.** The cavity Q-factor decreases as rather high dissipative currents appear in a sample, which causes a non-resonant absorption of the microwave energy.

**Problem 2.** The resonance line is distorted: the symmetric Lorentzian line becomes asymmetric with a Dyson shape. This occurs as the skin-layer depth $\delta$ becomes sufficiently small [5]:

$$\delta < 4d, \quad (2)$$

where $d$ is the sample thickness.

**Problem 3.** The skin-effect causes screening: the microwave magnetic field cannot penetrate across the whole sample depth, being concentrated in a certain effective volume $V_{eff}$ that is smaller than the geometric volume $V$. This nonuniformity of the microwave distribution is manifested when

$$\delta < d. \quad (3)$$

**Problem 4.** Correct integration of a resonance line with the Dyson shape is necessary for comparison with the Lorentzian line shape of the reference sample.

Our study is concerned with the procedure for taking these specific features into account. Experiments were performed on n-Ga:As samples cut-out from a single ingot with an initial impurity concentration of $3.6 \times 10^{17}$ cm$^{-3}$ and compensated by neutron transmutation doping [6]. Parameters of the samples under study were reported in [7, 8]. The results of the present study enable a rather accurate determination of the spin density in conducting samples near the IM phase transition at various impurity center concentrations, degrees of compensation, and sample temperatures.

## 2 VARIATION OF THE CAVITY Q-FACTOR (PROBLEM 1)

To solve the problem of taking into account the variation of the cavity Q-factor with the sample conductivity, we used a rectangular cavity of the $TE_{103}$ type (from the set of a Varian ESR spectrometer) which has two magnetic-field antinodes. The quartz part of an ESR-910 cryostat with a sample was placed in one of these antinodes. In the other, we placed a reference sample from the same set, which had the form of a narrow tube 5 mm in diameter, with a rated linear spin density of $2.58 \times 10^{15}$ spins/cm. The positions of the reference and sample in the $TE_{103}$ cavity are shown in Fig. 1. It should be noted that, to a first approximation, the microwave magnetic field strength is the same in both antinodes, but a certain difference may arise because a noticeable amount of quartz with a dielectric constant of 3.75, of which the cryostat is made, is present in one of the antinodes.

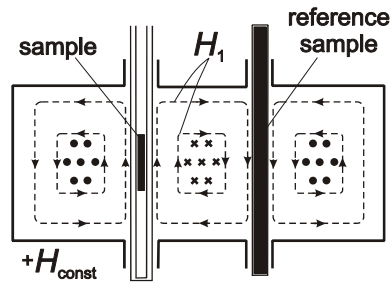

Fig. 1. Positions of the sample and reference in the antinodes of the microwave magnetic field in the $TE_{103}$ cavity.

With the cavity size and the configuration of the microwave magnetic field taken into account, the effective spin number of reference sample manifested in the resonance had the value specified above. The reference was at room temperature, which ruled out any direct influence of temperature on the amplitude of its ESR signal ($A$). This amplitude was only affected by the variation of the cavity $Q$-factor due to the temperature dependence of the sample conductivity:

$$A(T) \propto Q(T), \quad (4)$$

where $T$ is the sample temperature.

Thus, the relative change in the cavity Q-factor is accounted for by the ratio of signal amplitudes of reference sample at different sample temperatures:

$$Q(T_1)/Q(T_2) = A(T_1)/A(T_2), \quad (5)$$

where $T_1$ and $T_2$ are the sample temperatures, and $Q(T_1)$ and $Q(T_2)$, the cavity Q-factors at these temperatures.

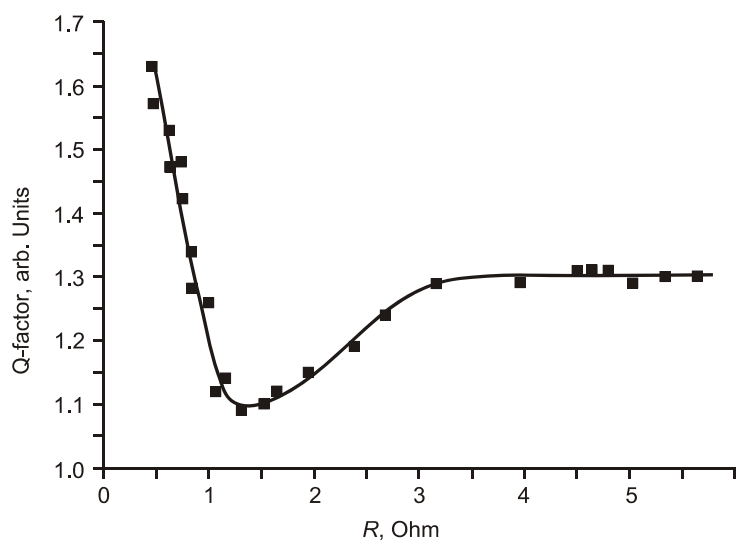

Fig. 2. Cavity Q-factor vs. the sample resistance.

The sample was placed in the same cavity as the reference and the amplitude of its signal corresponded to the same Q-factor. The variation of the cavity Q-factor with the sample resistance is shown in Fig. 2 for a particular sample no. 9 from the set under study [7, 8]. It can be seen that near the insulating to metallic state transition, the Q-factor starts to fall because the sample resistance decreases. This occurs, beginning at a certain resistance $R_0$, and continues until the depth of the skin layer becomes equal to the sample thickness. As the resistance decreases further, the cavity Q-factor starts to grow because the active volume of the sample decreases when the skin-layer thickness falls.

## 3 NONUNIFORMITY OF THE MICROWAVE FIELD AND DISTORTION OF THE LINE SHAPE IN CONDUCTING SAMPLES (PROBLEMS 2 AND 3)

The nonuniformity of the microwave field distribution in a sample appears when the sample thickness becomes comparable with the skin-layer depth determined by the resistance. In lightly doped samples at measurement temperatures below 10 K, this depth substantially exceeds the sample thickness and has nearly no effect on the ESR spectrum. In those samples at higher temperatures, when spins appear in the conduction band, the ESR signal disappears because of the sharp decrease in the relaxation time [4]. A different behavior is observed for more heavily doped samples near the IM phase transition. The region in which the ESR signal is observed extends toward higher temperatures up to 100 K, when samples already have a noticeable conductivity. This behavior coincides with the ESR in the metals [9, 10].

It is necessary to take into account in such a wide temperature range that the sample resistance markedly decreases with increasing temperature. Let us analyze how the ESR signal varies as the semiconductor resistance decreases in the temperature range from 2 to 100 K. The experimentally measured resistivity of the samples under study is presented in Fig. 3.[1]

It is noteworthy that the skin-layer depth at a frequency of 10 GHz ($\lambda \approx 3$ cm) is given by the known relation

$$\delta = (c^2/2\pi\sigma\omega)^{1/2} = 0.503\rho^{1/2}, \qquad (6)$$

where $\rho = 1/\sigma$ ($\sigma$ is the conductivity of the semiconductor) is measured in ohm centimeters, and $\delta$, in millimeters.

We used in our measurements $1\times3\times10$ mm$^3$ samples. Hence follows that the microwave field penetrates across the whole thickness of a sample at its resistivity higher than, or comparable with, 1 ohm cm. That is why, a horizontal line M is drawn in Fig. 3, below which the effective sample volume should be introduced because the field already does not penetrate across the whole sample depth. It can be seen that for samples 1–4, which are the closest to the IM transition, this should be done in the whole temperature range. For higher resistivity samples, this effect should be taken into account at sufficiently high temperatures $T > 30$ K. For sample 14, which is the farthest, toward the insulator, from the IM transition, the skin-effect is not manifested at all.

An additional physical phenomenon associated with the skin-layer thickness consists in the distortion of the ESR line shape, described by the Dyson theory [5]. This theory believes that the electrons diffuse like free particles and electron magnetic moments are treated like free-particles moments. These electrons feel the microwave phase and amplitude what change in the sample volume. The resonance line distortion takes place due to magnetic moment interaction with that wave.

This theory is applicable up to a $d/\delta \leq 4$ ratio between the sample thickness $d$ and the skin-layer thickness $\delta$. At a sample thickness of 1 mm, this condition is satisfied at $\rho < 16$ ohm cm. This corresponds to the region below the characteristic line N in Fig. 3, into which most of curves fall. For higher resistivity samples, the absorption signal must have a Lorentzian shape.

Let us consider how the distortion of the Lorentzian line is manifested at resistivities $\rho < 16$ ohm cm. Figure 4 shows as an example the variation of the resonance absorption line for sample 10 between temperatures of 2.6 and 4 K. It can be seen that the line becomes asymmetric with increasing temperature. The asymmetry consists in that the high-field wing *B* of the derivative of the resonance absorption is broadened and its amplitude decreases, compared with the width and amplitude of the low-field wing *A*: $A_2/B_2 > A_1/B_1 > 1$. According to Feher and Kip's conclusions [3, 9], the *A/B* ratio can serve as a measure of distortion of the ESR line for conducting samples. Let us find the extent to which the experimentally obtained distorted resonance lines are in agreement with the Dyson theory [5].

---

[1] The authors are grateful to L.N. Ionov for measurements of the temperature dependences of the sample resistivity.

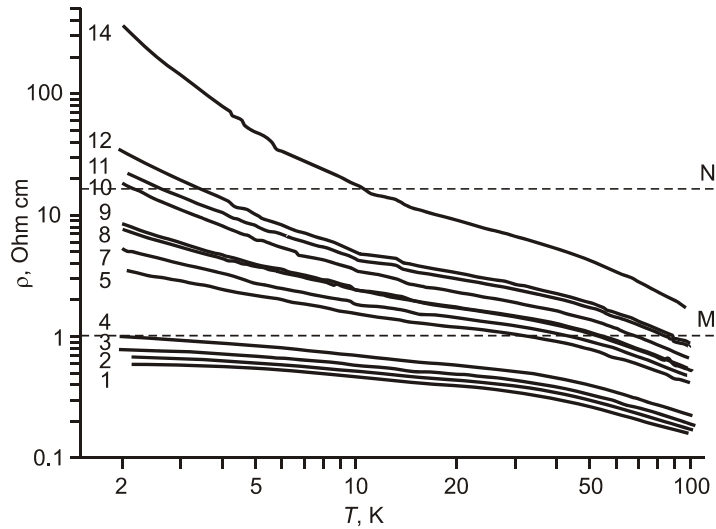

Fig. 3. Temperature dependence of the resistivity of samples with various spin densities.

We analyze the behavior of the *A/B* ratio on the basis of Feher and Kip's calculations made in terms of the Dyson model. Their results are presented in Fig. 5, in which the *A/B* ratio is plotted along the abscissa axis, and the $(T_D/T_2)^{1/2}$ ratio appearing as a parameter in the Dyson theory, along the ordinate axis. (Here, $T_D$ is the time of spin diffusion across the skin layer, and $T_2$, spin relaxation time.) It can be seen in this figure that, in the limit of large ratios between the time of spin diffusion across the skin layer ($T_D$) and the spin relaxation time ($T_2$), which is shown below to be the case in our study, $A/B = 2.7$ and remains constant until $(T_D/T_2)^{1/2}$ becomes on the order of unity. At $0.02 < (T_D/T_2)^{1/2} < 2$, $2.7 < A/B < 19$, with the last value being the limit for small *A/B* ratios.

Feher and Kip's calculations [3, 9] were made for metals. For semiconductors, we can introduce an additional restriction on this ratio. Indeed, it can be seen in Fig. 4 that the spin relaxation time $T_2$ is on the order of $10^{-8}$ s. This value can be calculated from the resonance Lorentzian line width ΔH:

$$T_2^{-1} = g\beta\Delta H/h, \qquad (6a)$$

where h is the Planck constant; β, Bohr's magneton.

This correlation usually satisfies for the Lorentzian line in semiconductors [10].

At the same time, the time of diffusion across the skin layer, $T_D = \delta^2/D$, where $D$ is the diffusion coefficient. At low temperatures, $D \approx 1$ cm$^2$/s and $\delta$ is on the order of 0.1 cm, i.e., $T_D \approx 10^{-2}$ s. Hence follows that, in the temperature range of our interest, the condition $(T_D/T_2)^{1/2} \gg 1$ is satisfied with a large safety margin.

Comparing the aforesaid with Fig. 6, we can see that, at low temperatures and most heavily doped samples (curves 1 and 2), this ratio corresponds to the case in which $T_D \gg T_2$, $A/B \approx 2.7$, i.e., the Dyson theory is valid for these samples. As the spin concentration decreases due to compensation, $T_D/T_2$ first somewhat increases and then gradually decreases to unity.

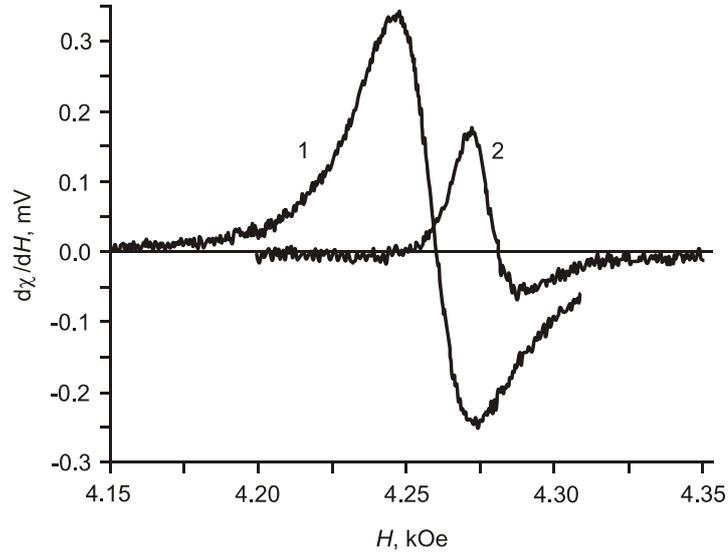

Fig. 4. ESR spectra of sample 10 at temperatures $T =$ (1) 2.6 and (2) 4 K.

For samples with high resistivity at low temperature, $A/B \to 1$, i.e., the Dyson line gradually becomes Lorentzian as the sample resistance increases. In this transition region, the Dyson theory is not valid because it shows that the wing ratio cannot be smaller than 2.7. With decreasing temperature, the line shape tends to become Lorentzian. The coincidence with this shape is better for the low-field part of the line. Figure 7 shows as an example the low-field part of the line for sample 10 at a temperature $T = 2$ K and the corresponding Lorentzian line. It can be seen that the coincidence is very good.

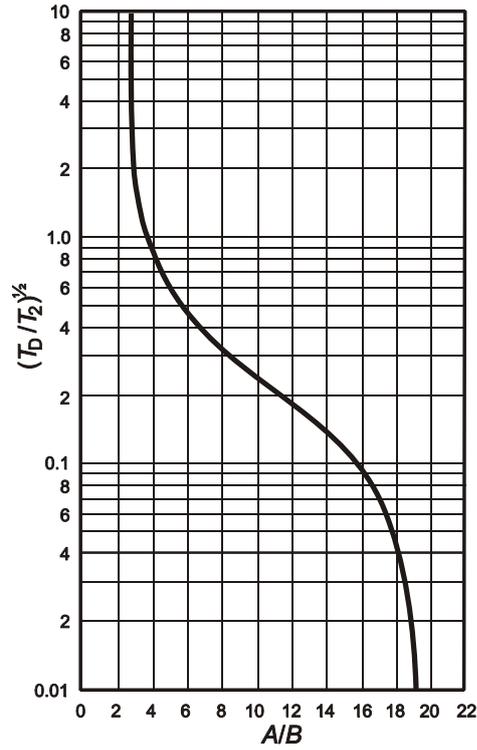

Fig. 5. Relationship between the ratio $A/B$ of the wings of the ESR line derivative for a conducting plate and the $(T_D/T_2)^{1/2}$ ratio [9].

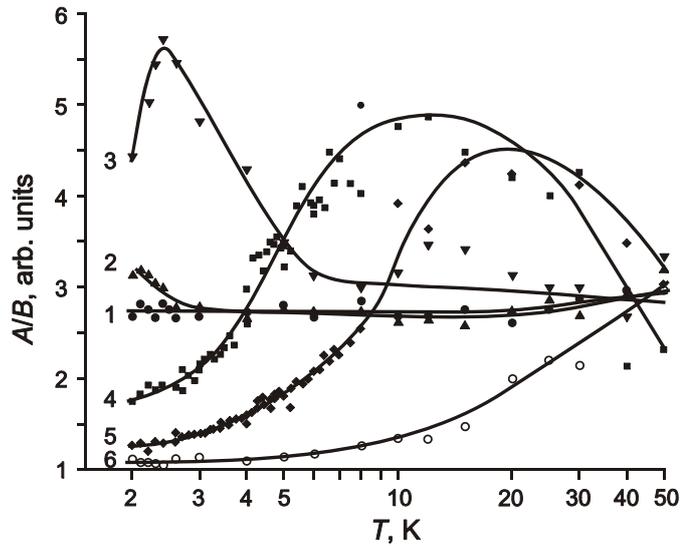

Fig. 6. Temperature dependences of the $A/B$ ratio (measure of distortion of the Lorentzian line) for several samples: (1) no. 1, (2) no. 3, (3) no. 7, (4) no. 9, (5) no. 10, and (6) no. 14.

Thus, we revealed the basic parameters of the experimentally obtained distorted resonance lines determined by the high conductivity of Ge:As samples. On this basis, we discuss in the next paragraph the way in which the second integral of the Dyson line can be calculated. This is necessary for correct determination of the density of the resonating spins.

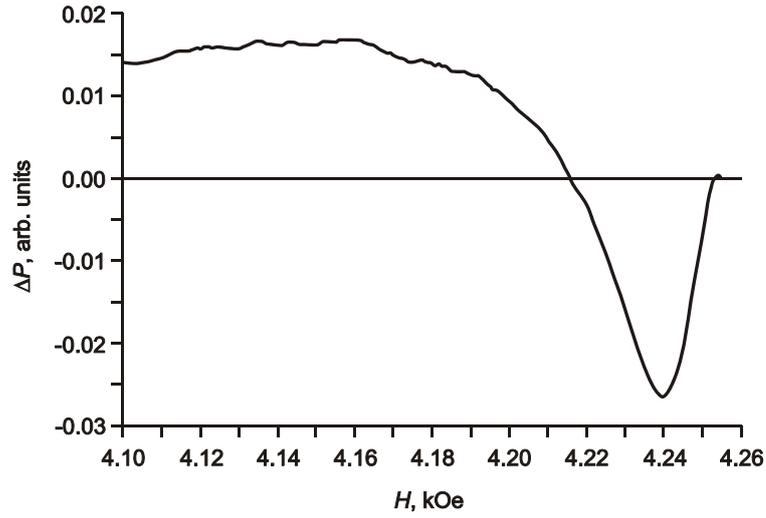

Fig. 7. Difference between the Lorentzian and experimentally obtained low-field part of the ESR line for sample 10, $T = 2$ K. It can be seen that this difference does not exceed 3% anywhere.

## 4 CORRECT INTEGRATION OF THE DYSON LINE (PROBLEM 4)

The least studied among the problems considered here is that of comparison of the Lorentzian and Dyson line shapes. The problem is associated with determining the magnetic susceptibility of a sample by integration of the absorption line, with a reference used. In our case, the role of reference was played by a certified sample from Varian company. Its resonance absorption line has a Lorentzian shape. As shown above, the same shape is observed for sufficiently high-resistivity samples, far from the IM phase transition. At the same time, low-resistivity samples have a Dyson line. We perform integration for these samples by using Feher and Kip's results for metals [3, 9].

The shape of a Lorentzian line is described by the expression:

$$P = \omega H_1^2 \delta S \omega_0 \chi_0 T_2 / 4[1 + T_2^2(\omega-\omega_0)^2], \qquad (7)$$

where $\omega$ and $\omega_0$ are the frequency of the microwave field and the resonance frequencies, respectively; $H_1$, amplitude of the microwave field; $\delta = (c^2/2\pi\sigma\omega)^{1/2}$, depth of the skin layer of the sample ($\sigma$ is the conductivity of the semiconductor); $S$, surface area of the semiconductor sample; $\chi_0$, magnetic susceptibility of the sample; and $T_2$, time of spin relaxation in the sample.

We use the resonance relation for ESR:

$$\omega = (g\beta H/\hbar) = \gamma H, \qquad (8)$$

where $\hbar$ is the Planck constant; $\beta$, Bohr's magneton; and $\gamma$, gyromagnetic ratio.

Hence we obtain an expression for the Lorentzian line shape as a function of the magnetic field $H$:

$$P = \gamma H H_1^2 \, \delta S \gamma H_0 \chi_0 T_2/4[1 + T_2^2\gamma^2(H - H_0)^2], \qquad (9)$$

where $H_1$ is the amplitude of the microwave magnetic field, and $H_0$, resonance magnetic field.

Replacing the numerator in expression (7) with the maximum power $P_{max}$ absorbed at resonance, we obtain the shape of the Lorentzian line:

$$P_L = P_{max}/(1 + [(H - H_0)/(1/2)\Delta H_{1/2}]^2), \qquad (10)$$

where $\Delta H_{1/2}$ is the full width at half-maximum (FWHM) of the resonance line.

Using the Dyson theory, Feher and Kip [9] analyzed the shape of the resonance line for conducting samples and derived formulas describing this shape for a number of limiting cases. As already shown above, samples having low conductivity, compared with good metals, and a skin-layer thickness $\delta < 4d$ that is comparable with the sample thickness, are described in the approximation in which the time $T_T$ in which a spin traverses the whole sample thickness is on the order of, or longer than the time $T_D$ of diffusion across the skin layer. In addition, $T_T$, $T_D \gg T_2$. It is noteworthy that the transition from the Lorentzian to Dyson line occurs in that range of conductivities in which the electromagnetic field penetrates across the whole sample depth.

Using these approximations, Feher and Kip derived for the Dyson shape of the resonance absorption line the following expression:

$$P_D = P_L/2 - P_L[(H - H_0)/(1/2)\Delta H_{1/2}]/2 = P_L[1 - (H - H_0)/(1/2)\Delta H_{1/2}]/2 \qquad (11)$$

Formulas (10) and (11) for the Lorentzian and Dyson line shapes are compared in Fig. 8.

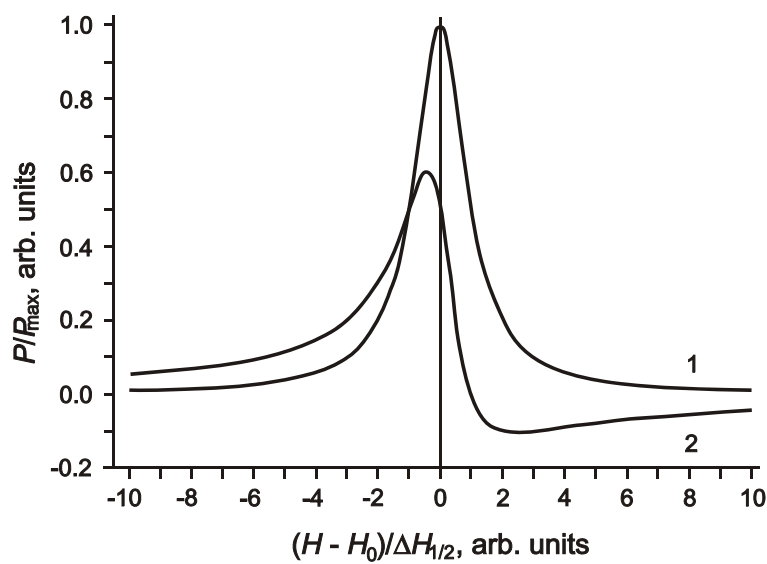

Fig. 8. Comparison of (1) Lorentzian and (2) Dyson shapes of resonance absorption lines.

It can be seen in the figure that, at the same values of the parameters, the amplitude of the Dyson line at its maximum is only 0.6 of the amplitude of the Lorentzian line. In addition, the maximum of the Dyson line is shifted downfield by $0.4[(H − H_0)/(1/2)\Delta H_{1/2}]$.

To accurately determine the paramagnetic susceptibility and spin density, it is necessary to compare integrals of absorption lines for the sample under study and a control. As the control sample served a calibrated sample from Varian company ($2.58 \times 10^{15}$ spins/cm). A test demonstrated that the line from this sample is Lorentzian with high precision. The Dyson line from a conducting sample should be compared with that from the control.

For comparison, we calculated integrals of the Lorentzian and Dyson lines by using formulas (10) and (11). It was found that the integral of the Lorentzian line from −10 to +10 is 2.94676. The integral of the Dyson line within the same limits is 1.47338, i.e., it is two times smaller. However, the error can be reduced as follows.

Modern ESR spectrometers record the derivative of the absorption with respect to the magnetic field, $dP/dH$, and the magnetic field at which the derivative of the absorption line changes its sign is rather easily determined. We used this field as the upper integration limit. As expected, the integral for the Lorentzian line was found in this case to be half the full integral (1.47338).

For the derivative of the Dyson line, the sign-change point is shifted to weaker fields and lies at a point −0.4 from the center of the Lorentzian line. The integral within the limits from −10 to −0.4 was found to be 1.6637. Thus, twice integrating the low-field part of the derivative of the absorption line, we obtain for the Dyson line an integral that differs from the integral for the Lorentzian line by 13%. When the spin density is determined from results of ESR measurements, it is commonly considered that the method has an error of about 20%, and the technique we suggest for data processing meets this requirement.

It is clear that the difference of integrals of the Dyson and Lorentzian lines depends on the integration region. If we plot the difference of these integrals as it is done in Fig. 9, it is found that there is an upper integration limit at which these integrals are equal. However, this point has no specific features and it is rather difficult to determine its position. Therefore, it is much more convenient to use the procedure suggested above by integrating up to the field at which the derivative of the absorption line changes its sign.

It follows from Fig. 6 that, as the sample resistance increases in a certain temperature range, the *A/B* ratio grows because of the decrease in the negative part of the derivative of the signal. However, the ratio between the integral of the positive parts of the derivatives of resonance absorption lines remains approximately the same. In numerical integration of distorted lines in this temperature range, no specific features associated with changes in the *A/B* ratio were observed.

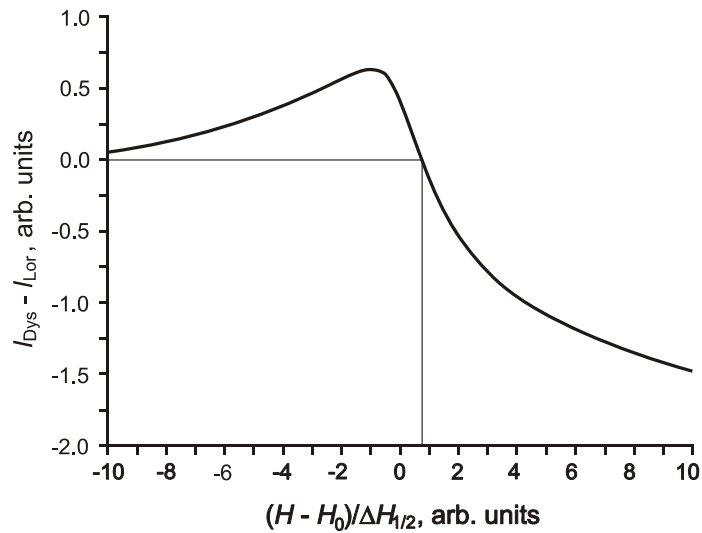

Fig. 9. Difference between the integrals of the Dyson and Lorentzian lines in integration from $-10$ to values on the X axis. It can be seen that the integrals coincide at an upper integration limit that is somewhat smaller than unity.

As already noted, further decrease in temperature makes smaller the integration error and, as it can be seen in Fig. 7, the line in the most lightly doped samples becomes Lorentzian and its integral adequately reflects the magnetic susceptibility and, consequently, the spin density.

Thus, the magnetic susceptibility can be determined in the case of Dyson lines by comparing the second integrals of the positive parts of derivatives of this resonance absorption line and a line of the Lorentzian shape. The error of this procedure does not exceed 10–15%.

**5 CONCLUSIONS**

A procedure for determining the magnetic susceptibility described for the example of Ge:As. The procedure includes three parts.

(1) Use of an ESR cavity with two magnetic field antinodes, with a sample under study placed in one of these, and a reference sample into the other.

(2) Measurement of the temperature dependence of the resistivity of a sample ("poor" conductor) in order to take into account the nonuniformity of the microwave field distribution in the sample.

(3) Double integration of the measured positive part of the derivative of the Dyson resonance line, followed by application of formula (11). The procedure yields the magnetic susceptibility with an error not exceeding 15%.

**Acknowledgments** The authors are grateful for financial support to the Russian Foundation for Basic Research (grant no. 10-02-00629), RF Ministry of education and science (RF Presidential grant NSh-3008.2012.2), Presidium and Department of physical sciences of the Russian Academy of Sciences, and EC Research Executive Agency ("People" Programme, proposal 295180).